\documentclass[a4paper,10pt]{article}

\usepackage[margin=1in]{geometry}
\usepackage{graphicx}
\usepackage{hyperref}

\title{Anomalous Spin Precession Frequency Analysis in the Muon $g-2$ Experiment at Fermilab\\ \vspace{1em} \large Contribution to the 25th International Workshop on Neutrinos from Accelerators}
\author{On Kim on behalf of the Muon $g-2$ Collaboration\thanks{\url{https://muon-g-2.fnal.gov/collaboration.html}}}
\date{}

\begin{document}
\maketitle

\begin{abstract}
The Muon $g-2$ experiment at Fermilab aims to measure the muon anomalous magnetic moment with an unprecedented precision of 140 parts per billion (ppb). Data collection concluded in June 2023, and analysis of the largest dataset (2021–2023) is underway. Previous publications based on data from 2018–2020 established the experimental foundation. This document provides an overview of the measurement of the muon anomalous spin precession frequency ($\omega_a$) and the associated systematic corrections. The precision of these results directly tests the Standard Model’s completeness, making the experiment a cornerstone in the field of particle physics.
\end{abstract}

\section{Introduction}
\begin{figure}[htb]
    \centering
    \includegraphics[width=0.65\textwidth]{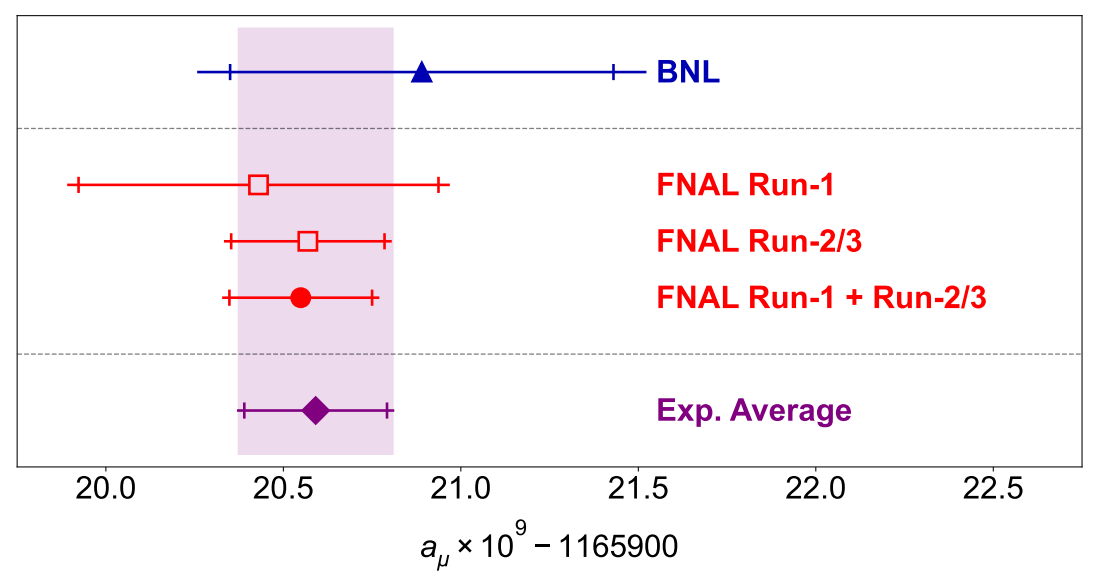}
    \caption{The magnetic anomaly of the muon ($a_\mu$) measured at BNL (blue triangle) and Fermilab (red points). The Fermilab results dominate the experimental average~\cite{Aguillard2023}. Tick marks indicate statistical uncertainties. This paper reports on the latest measurement using Run-2/3 data. Adapted from Ref.~\cite{Aguillard2023}.}
    \label{fig:amu}
\end{figure}

The Muon $g-2$ experiment at Fermi National Accelerator Laboratory (FNAL) seeks to determine the muon magnetic anomaly ($a_\mu$) with a precision of 140 ppb. The first result, based on 2018 data (Run-1), confirmed a previous measurement performed at Brookhaven National Laboratory (BNL), with the experimental average deviating from the Standard Model prediction by 4.2 standard deviations~\cite{Abi2021}. Subsequent results from 2019–2020 data (Run-2/3), published in August 2023, halved the uncertainty while maintaining consistency with prior measurements~\cite{Aguillard2023}, as shown in Fig.~\ref{fig:amu}.

The anomaly is derived from the formula:
\begin{equation}
    a_\mu = \frac{\omega_a}{\tilde{\omega}_p'(T_r)} \frac{\mu_p'(T_r)}{\mu_e(H)} \frac{\mu_e(H)}{\mu_e} \frac{m_\mu}{m_e} \frac{g_e}{2},
\end{equation}
where $\mathcal{R}'_\mu \equiv \omega_a / \tilde{\omega}_p'(T_r)$ is the only experimentally measured term. The remaining terms, such as $\mu_p'(T_r)$ (proton magnetic moment at reference temperature $T_r$), $\mu_e(H)$ (electron magnetic moment in the hydrogen bound-state), $m_{\mu,e}$ (muon and electron masses), and $g_e$ (electron $g$-factor), are constants with a combined uncertainty of 25 ppb. The anomalous spin precession frequency of the muon, $\omega_a$, depends on the magnetic field strength $B$ in the storage ring:
\begin{equation} \label{eq:omega_a}
    \omega_a = a_\mu \frac{q}{m} B.
\end{equation}

To measure $a_\mu$, both $\omega_a$ and the magnetic field must be determined. The magnetic field is quantified via the proton Larmor precession frequency ($\omega_p$), described in detail in Ref.~\cite{Albahri2021Field}. The next section outlines the measurement of $\omega_a$.

\section{Anomalous Spin Precession Frequency Analysis}
Muon decay provides a natural way to analyze the muon spin due to parity-violating weak interactions, with high-energy decay positrons preferentially emitted along the muon spin direction. This property is fundamental to the experiment, as it allows the anomalous spin precession frequency of the muon to be directly inferred from the distribution of decay positrons. The number of detected positrons above a threshold energy oscillates at the anomalous spin precession frequency, with their energies measured by 24 calorimeter systems positioned around the storage ring. This oscillatory pattern serves as a key observable for extracting $\omega_a$, the anomalous spin precession frequency, and encapsulates the underlying physics of the experiment.

The oscillatory exponential decay (referred to as the ``wiggle plot") provides the raw data for $\omega_a$ extraction. This characteristic oscillation, observed over the muon lifetime, is modeled to extract $\omega_a^m$, the uncorrected precession frequency. Beam dynamics introduce corrections to this raw measurement, as discussed in Ref.~\cite{Albahri2021BD}. These corrections are necessary to account for systematic effects arising from non-ideal beam motions, ensuring that $\omega_a$ accurately represents the muon's interaction with the magnetic field. The fit function used to model the wiggle plot is~\cite{Albahri2021OmegaA}:
\begin{equation}
    N(t) = N_0 e^{-t/\tau} \cdot \eta_N(t) \cdot \{ 1 + A \cdot \eta_A(t) \cos(\omega_a^m t + \phi_0 + \eta_\phi(t)) \},
\end{equation}
where $\eta_N(t)$, $\eta_A(t)$, and $\eta_\phi(t)$ represent time-dependent acceptance corrections arising from beam dynamics. These functions, which are empirically determined, correct for variations such as changes in the muon distribution or detector response over time~\cite{Albahri2021OmegaA}.

\begin{figure}[t]
    \centering
    \includegraphics[width=0.65\textwidth]{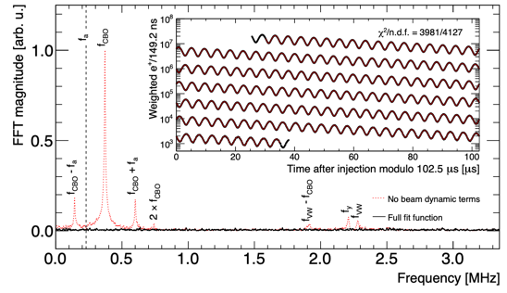}
    \caption{Residual FFT spectrum for the wiggle plot fit. Red dashed lines represent fits without beam dynamics corrections ($\eta(t)$), while black solid lines include them. The inset shows decay positron counts over time with the fit overlaid. Adapted from Ref.~\cite{Aguillard2023}.}
    \label{fig:wiggleFFT}
\end{figure}

Without these corrections, significant residual peaks appear in the Fourier transform (FFT) of the fit residuals, as shown in Fig.~\ref{fig:wiggleFFT}. The most prominent peak corresponds to Coherent Betatron Oscillation (CBO), a collective radial oscillation of the muon beam around its equilibrium position. This effect introduces a systematic bias in $\omega_a$ because the oscillation modulates both the spatial distribution of muons and their decay positrons. Vertical oscillations, though less pronounced than CBO, also contribute to these distortions. The acceptance correction terms $\eta_N(t)$, $\eta_A(t)$, and $\eta_\phi(t)$ are carefully constructed to suppress these artifacts, thereby improving the accuracy of $\omega_a$ extraction.

Beyond the corrections, the wiggle plot analysis is sensitive to other factors that can affect the accuracy of $\omega_a$. These include pileup effects, where multiple positron signals overlap within a short time interval, and gain instabilities in the calorimeter system, which can alter the energy calibration. The analysis framework incorporates these corrections iteratively, ensuring a robust and systematic approach to extracting $\omega_a$.

\begin{figure}[htb]
    \centering
    \includegraphics[width=0.65\textwidth]{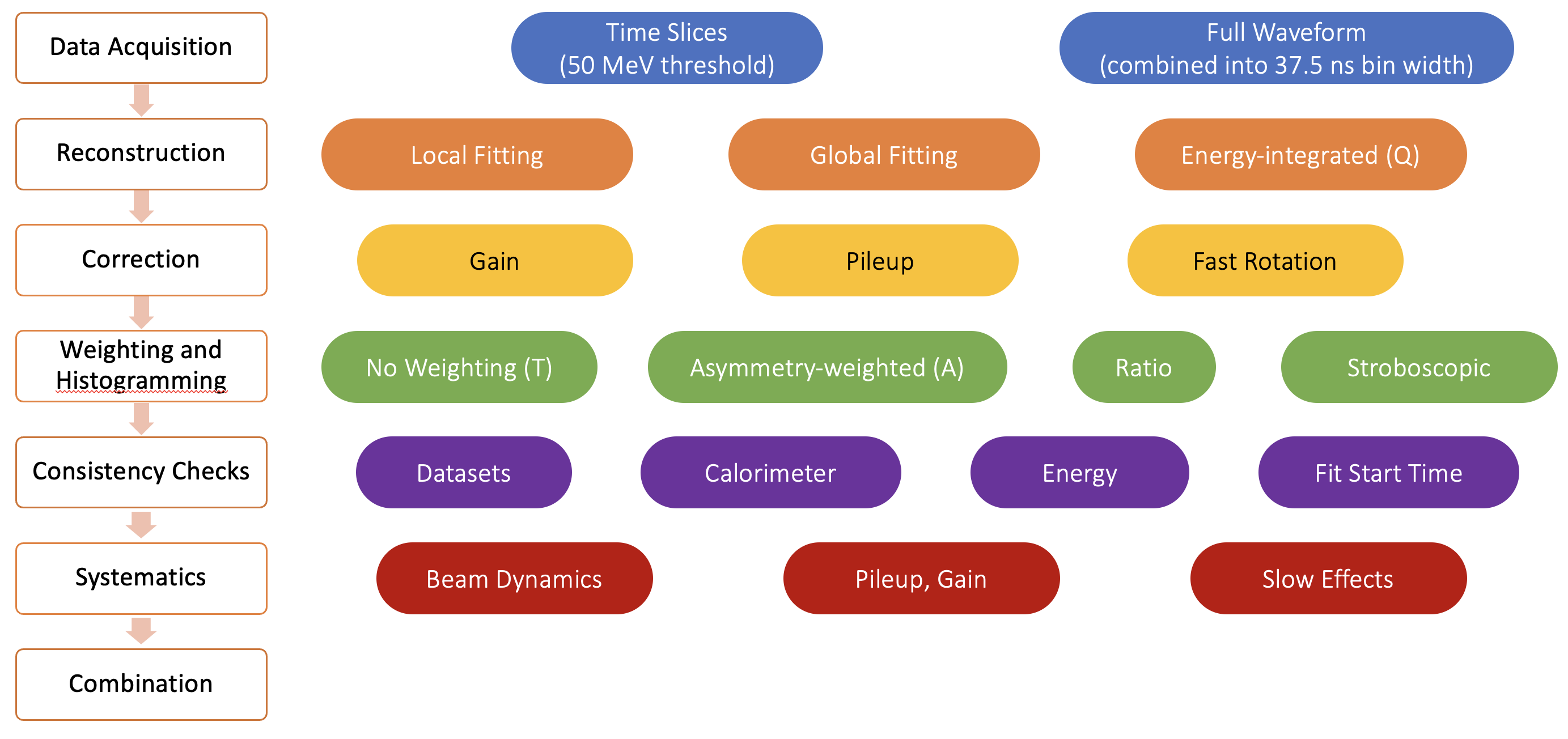}
    \caption{Analysis flow for $\omega_a$ extraction and systematics assessment. The process integrates raw data modeling, beam dynamics corrections, and iterative systematic studies to ensure the accuracy and reliability of the extracted precession frequency.}
    \label{fig:anachain}
\end{figure}

The analysis flow for $\omega_a$ extraction and systematic assessment is illustrated in Fig.~\ref{fig:anachain}. This flowchart highlights the interplay between raw data, corrections, and systematic studies, emphasizing the multi-step nature of the analysis. Each stage is designed to minimize both statistical and systematic uncertainties, leveraging the high statistics and precision of the Muon $g-2$ dataset.

Through extensive systematic analysis, we achieved a 25 ppb systematic uncertainty on the measured anomalous spin precession frequency using Run-2/3 data, as reported in our second scientific results~\cite{Aguillard2023, Aguillard2024PRD}.

\section{Key Methodologies and Corrections}
\begin{enumerate}
    \item \textit{DAQ System:}  
    The system produces two data streams:
    \begin{itemize}
        \item An event-based stream extracts 40 ns windows around signals exceeding $\sim$50 MeV.
        \item A continuous stream samples waveforms for an integrated energy analysis.
    \end{itemize}

    \item \textit{Reconstruction Approaches:}  
    Event-based reconstruction, utilizing the first DAQ data stream, employs the following methods:
    \begin{itemize}
        \item Local-fitting: Fits individual crystal waveforms and clusters hits into positron candidates. Each calorimeter consists of a $6 \times 9$ array of $\mathrm{PbF}_2$ Cherenkov crystals.
        \item Global-fitting: Simultaneously fits $3 \times 3$ crystal arrays to determine combined time and energy data.
    \end{itemize}
    Alternatively, an energy-based method, utilizing the second DAQ data stream, bypasses waveform fitting and directly analyzes deposited energy distributions, offering complementary insights.

    \item \textit{Corrections:}  
    \begin{itemize}
        \item Gain corrections address temperature fluctuations, beam-induced sag, and silicon multiplier (SiPM) deadtime. The SiPMs, attached to the end of each crystal, detect the Cherenkov light and digitize the signal for further processing.
        \item Pileup corrections mitigate unresolved close-in-time positron hits.
        \item Fast-rotation handling reduces distortions from cyclotron-period modulation.
    \end{itemize}
    These corrections are critical for achieving the experiment's stringent precision goals.

    \item \textit{Data-Weighting Schemes:}  
    \begin{itemize}
        \item Threshold (T): Equal weights for all positrons above a threshold.
        \item Asymmetry-weighted (A): Maximizes statistical power via energy-dependent asymmetry weights.
        \item Integrated-energy (Q): Uses energy weights without resolving individual positrons.
    \end{itemize}
    A ratio method further mitigates slow variations by histogramming time-shifted subsets.

    \item \textit{Robustness Checks:}  
    Stability was validated through:
    \begin{itemize}
        \item Start time scans (slow effects such as gain stability).
        \item Calorimeter scans (detector-specific variations).
        \item Energy scans (pileup and gain dependence).
    \end{itemize}
    These checks ensure that the extracted $\omega_a$ values are resilient to systematic influences.

    \item \textit{Systematics:}  
    Key sources of systematic uncertainty include:
    \begin{itemize}
        \item CBO distortions (largest contributor, accounting for 21 ppb of the total 25 ppb uncertainty in the measured $\omega_a^m$ analysis~\cite{Aguillard2024PRD}).
        \item Pileup and gain corrections.
        \item Residual slow effects such as energy scale variation.
    \end{itemize}
    The total systematic uncertainty across datasets ranges from 24 to 31 ppb, reflecting rigorous control over potential biases.

    \item \textit{Combining Measurements:}  
    Averages from six analyses for each dataset (Run-2, Run-3a, Run-3b) provided the final $\omega_a$ values. Less precise methods were excluded to optimize the overall accuracy and reliability.
\end{enumerate}

\section{Conclusion}
The Muon $g-2$ experiment has reduced the statistical uncertainty on $\omega_a$ to 0.20 ppm using four times the Run-1 statistics. Systematic analysis achieved an uncertainty of 25 ppb, advancing the experiment toward its precision goal. These achievements mark a significant step forward in the effort to test the limits of the Standard Model and search for potential signs of new physics.

\section*{Acknowledgement}
O.~K. is supported by the U.S. DOE under contract no. DE-SC0021616.
This manuscript has been authored by Fermi Research Alliance, LLC under Contract No. DE-AC02-07CH11359 with the U.S. Department of Energy, Office of Science, Office of High Energy Physics.

\bibliographystyle{JHEP}
\bibliography{references}

\end{document}